\documentclass[aps,prd,preprint,superscriptaddress,preprintnumbers,nofootinbib]{revtex4-1}

\usepackage{setspace}
\usepackage{braket}
\usepackage{amsmath}
\usepackage{amsfonts} 
\usepackage{amssymb}
\usepackage{graphicx}
\usepackage{hyperref}
\usepackage{tensor}
\usepackage{siunitx}
\usepackage{physics}

\usepackage{appendix}
\newcommand{\be}{\begin{equation}}
\newcommand{\ee}{\end{equation}}

\newcommand{\beq}{\begin{equation}}  \newcommand{\eeq}{\end{equation}}
\newcommand{\beqn}{\begin{eqnarray}}
 \newcommand{\eeqn}{\end{eqnarray}}
\newcommand{\bal}{\begin{aligned}}   \newcommand{\eal}{\end{aligned}}
\newcommand{\bea}{\begin{eqnarray}}  \newcommand{\eea}{\end{eqnarray}}

\newcommand\blfootnote[1]{%
  \begingroup
  \renewcommand\thefootnote{}\footnote{#1}%
  \addtocounter{footnote}{-1}%
  \endgroup
}

\hypersetup{colorlinks=true,
			urlcolor=black,
			citecolor=blue,
			linkcolor=magenta,}

\begin{document}
\count\footins = 1000 

\blfootnote{\href{mailto:niccolo.cribiori@tuwien.ac.at}{niccolo.cribiori@tuwien.ac.at}\\ \href{mailto:dieter.luest@lmu.de}{dieter.luest@lmu.de}\\ \href{mailto:mscalisi@mpp.mpg.de}{mscalisi@mpp.mpg.de}}

\begin{flushright}
LMU-ASC 10/21\\
MPP-2021-62
\end{flushright}

\title{\Large   The Gravitino and the Swampland}

{~}

\author{\vspace{0.5cm}
\large Niccol\`o Cribiori\vspace{0.5cm}}
 \affiliation{\setstretch{1.25}Institute for Theoretical Physics, TU Wien,  Wiedner Hauptstrasse 8-10/136, A-1040 Vienna, Austria\\[1.5ex]}
\author{\large Dieter~L\"ust}
 \affiliation{\setstretch{1.25}Arnold-Sommerfeld-Center for Theoretical Physics, Ludwig-Maximilians-Universit\"at, 80333 M\"unchen, Germany\\[1.5ex]}
\affiliation{\setstretch{1.25}Max-Planck-Institut f\"ur Physik (Werner-Heisenberg-Institut),
             F\"ohringer Ring 6,
             80805, M\"unchen, Germany\\[1.5ex]\vspace{0.8cm}}
\author{\large Marco Scalisi}
\affiliation{\setstretch{1.25}Max-Planck-Institut f\"ur Physik (Werner-Heisenberg-Institut),
             F\"ohringer Ring 6,
             80805, M\"unchen, Germany\\[1.5ex]\vspace{0.8cm}}

\begin{abstract}
\vspace{.5cm}
\noindent \setstretch{1.25}
We propose a new swampland conjecture stating that the limit of vanishing gravitino mass corresponds to the massless limit of an infinite tower of states and to the consequent breakdown of the effective field theory. We test our proposal in large classes of models coming from compactification of string theory to four dimensions, where we identify the Kaluza-Klein nature of the tower of states becoming light. We point out a general relation between the gravitino mass and abelian gauge coupling in models with extended supersymmetry, which can survive also in examples with minimal supersymmetry. This allows us to connect our conjecture to other well established swampland conjectures, such as the weak gravity conjecture or the absence of global symmetries in quantum gravity. We discuss phenomenological implications of our conjecture in (quasi-)de Sitter backgrounds and extract a lower bound for the gravitino mass in terms of the Hubble parameter.

\end{abstract}

\maketitle
\thispagestyle{empty}

{\setstretch{1.25}
\hypersetup{linkcolor=black}
\tableofcontents}

\section{Introduction}\label{intro}

Supersymmetry (SUSY) has been elusive so far. The LHC has probed energies up to \mbox{13 TeV} and found no evidence of supersymmetric particles. At present, given the observed small value of the dark energy density, an estimation for the scale $M_{SUSY}$ at which we expect supersymmetry to be broken is given by the mass of the {\it gravitino} $m_{3/2}$ as
\be
M_{SUSY}^2 \simeq m_{3/2}\ M_P \,,
\ee
with $M_P$ being the reduced Planck mass. The gravitino is in fact a key-particle in any effective field theory (EFT) coming from supergravity and string theory. However, so far, theoretical arguments have struggled to provide definite indication on its expected mass range, besides model-dependent results. Motivated by all of this, we would like to pose the following question: is there any fundamental property of quantum gravity that might give us information about the mass of the gravitino?

The swampland program \cite{Vafa:2005ui,Ooguri:2006in,Palti:2019pca,vanBeest:2021lhn} aims to capture universal features which have to be present in any consistent low energy effective theory of quantum gravity. Given our limited understanding of quantum gravity and string theory, these properties are at present formulated as conjectures. This approach is after all not that unique in the development of physics. For example, in the context of quantum mechanics, a number of `principles' were proposed to capture fundamental properties of the theory. This is the case with Heisenberg's uncertainty principle, or with Bohr's correspondence principle of the atomic model. A peculiarity of these principles is that they become trivial, once we send the Planck constant $h$ to zero.\footnote{It is temping to interpret in this sense also the celebrated Bekenstein--Hawking formula, $S = \frac{A}{4L_P^2}$, indicating that the entropy of a black-hole diverges in the classical limit $L_P\to 0$.} The same situation happens in the context of the swampland, where the statements lose their power once we go to the limit of infinite $M_P$.

Within the swampland program, a recurrent conclusion is that the limit of {\it vanishing mass} or {\it small gauge coupling}  typically signals a quantum gravity obstruction. Among the most studied conjectures, the Swampland Distance Conjecture (SDC) \cite{Ooguri:2006in} relates in fact an exponentially vanishing mass $m$ to an infinite geodesic distance $\Delta$ in field (moduli) space, such as
 \be
m = M_P\ e^{-\Delta } \,.
\label{dsc}
\ee 
This is the mass of an infinite tower of states. As $m$ becomes small, these states cannot be neglected and the effective description needs to be reformulated.  In the limit of vanishing mass, the cut-off of the theory experiences a similar drop-off and any possible effective gravitational description breaks down. The SDC is  supported by two generic situations in string theory, namely KK modes or string modes \cite{Lee:2019wij}.

Another well tested conjecture is the Weak Gravity Conjecture (WGC) \cite{ArkaniHamed:2006dz}, postulating a relation between masses and (abelian) gauge couplings. This conjecture can be understood as an obstruction in taking the continuous limit of vanishing gauge coupling, for a particle charged under an abelian gauge symmetry. Indeed, in that limit, a global symmetry would be restored, contradicting another well tested swampland principle: the absence of global symmetries in quantum gravity \cite{Banks:1988yz}.

More recently, obstructions to the limit of vanishing mass of various objects have been studied from different perspectives and related to the swampland program. For example, on backgrounds with positive energy, the consequences of the Higuchi bound \cite{Higuchi:1986py} for integer higher spin particles and the string tower have been analysed in \cite{Noumi:2019ohm,Lust:2019lmq,Scalisi:2019gfv,Luben:2020wim,Kato:2021rdz,Parmentier:2021nwz}, while \cite{Montero:2019ekk} proposes a general lower bound for the mass of any particle charged under an abelian symmetry. The specific case of massive spin-2 fields and of massive gravity has instead been considered in detail in \cite{Klaewer:2018yxi}. Fermions have maybe received less attention than bosons, even if considerations on their interactions can be used to deduce important information on the breaking of supersymmetry in low energy effective field theories \cite{Palti:2020tsy} and on the 
lightest neutrino in the Standard Model \cite{Gonzalo:2021fma}.

In this work, we focus exclusively on the spin-3/2 case corresponding to the gravitino, the superpartner of the graviton. Specifically, we suggest that the limit of vanishing gravitino mass should belong to the swampland, independently of the background. In fact, we conjecture that this limit is always accompanied by a infinite tower of states becoming massless, thus invalidating any possible effective description. We provide evidence coming from various examples in string compactifications with $\mathcal{N}=1,2$ supersymmetry in four dimensions, in anti-de Sitter, Minkowski and also de Sitter backgrounds.\footnote{Here, we assume the possibility that de Sitter constructions might be valid in string theory, although their existence has been challenged by a number of works in the literature, see for example \cite{Dvali:2013eja,Dvali:2014gua,Dvali:2017eba,Obied:2018sgi,Dvali:2018fqu,Ooguri:2018wrx,Dvali:2018jhn,Andriot:2018wzk,Garg:2018reu,Andriot:2018mav,Rudelius:2019cfh,Bedroya:2019snp,Blumenhagen:2020doa,Dvali:2020etd}.} 
Furthermore, we observe that, in extended supersymmetry, the gravitino mass and the abelian gauge coupling are related and thus the issue in having a small gravitino mass can be understood as an obstruction in restoring a global symmetry.  A similar mechanisms can be at work also in $\mathcal{N}=1$ models, in the case they have an underlying $\mathcal{N}=2$ supersymmetry, as we will show in a specific class of examples.

Interestingly, our proposal provides also a rationale behind some a priori unexpected correlations found in a series of previous works. In \cite{Palti:2020qlc}, it has been in fact argued that a vanishing $\mathcal{N}=1$ superpotential, corresponding to a vanishing gravitino mass, can be explained only in the presence of an underlying supersymmetry protection relating the model to a parent theory with a higher amount of preserved supercharges. In the absence of this specific mechanism, any allowed correction to the superpotential should in principle occur and thus the gravitino would be massive. In the framework of $\mathcal{N}=2$ supergravity in four dimensions, it has been noticed in \cite{Cribiori:2020use} a correlation between a vanishing gravitino mass and de Sitter extrema that are shown to clash with the magnetic WGC. In particular, $\det S_{AB} =0$ has been proposed as a swampland criterion on backgrounds with positive energy, $S_{AB}$ being the gravitino mass matrix.

Finally, since our conjecture is directly related to a drop-off of the quantum gravity cut-off, one can quite concretely investigate the phenomenological implications. Specifically, we focus on the case of inflation and show that it is possible to extract precise bounds on the gravitino mass, once we have information on the Hubble scale. We show that the strong form of our conjecture predicts that any detection of B-modes in the next-generation Cosmic Microwave Background (CMB) experiments \cite{Abazajian:2016yjj,Ade:2018sbj} would point at a lowest possible value of at least about $10^2$ GeV for the gravitino mass.
 
The rest of the paper is organized as follows. In section \ref{GMCsec}, we propose a new swampland conjecture on the mass of the gravitino. Then, we discuss the relation with the Anti-de Sitter Conjecture (ADC) \cite{Lust:2019zwm}. In section \ref{n1}, we test our proposal on various $\mathcal{N}=1$ models coming from string compactifications to four dimensions. In section \ref{n2}, we turn our attention to $\mathcal{N}=2$ models. We point out a relation between the mass and the abelian gauge coupling of the gravitino, which allows us to connect our conjecture to other well tested swampland proposals. In section \ref{Sec.QGcut-off}, we discuss how the quantum gravity cut-off is affected by a reduction of the gravitino mass. In section \ref{sec6}, we present the phenomenological implications of our conjecture in quasi-de Sitter space. We summarise our conclusions in section  \ref{sum}. Two appendices provide more technical details.

We would like to note that, while this work was under completion, the references \cite{Kolb:2021nob,Kolb:2021xfn,Dudas:2021njv,Terada:2021rtp} appeared, discussing aspects of gravitino physics in de Sitter backgrounds. Our work does not intend to contribute to this discussion, but rather presents an independent investigation.

\section{Gravitino Mass and Infinite Tower of States}
\label{GMCsec}
In this section, we first put forward a new swampland conjecture on the gravitino mass. Then we contrast it with the Anti-de Sitter Distance Conjecture, proposed in \cite{Lust:2019zwm}.

\subsection{The Gravitino Mass Conjecture}

The swampland program tells us that certain limits are forbidden within the validity regime of a given EFT. A primary and best studied example is the limit of infinite distances in field space, which is always accompanied by an infinite tower of light states \cite{Ooguri:2006in}. This is a typical UV effect, which ultimately leads to the breakdown of the EFT.

Observational evidence suggests that the mass of the gravitino might not be very small in the low-energy EFT describing our universe. This might be just an accident or, more intriguingly, it could constitute a universal property of effective theories consistent with quantum gravity.  Similarly to what described above, one may think that there should be a typical UV-obstruction in taking such a limit. In fact, in the following sections, we will show that a general feature of concrete string theory models is that the limit of small gravitino mass is always accompanied by a tower of Kaluza-Klein modes becoming light. Here, we would like to suggest that this is a universal feature of EFTs consistent with quantum gravity and formulate the following conjecture.

\begin{quote}

\noindent
\textbf{ Gravitino Mass Conjecture (GMC):}\\

\begin{spacing}{1.}
{\it The limit of small gravitino mass
\be\label{m32limit}
m_{3/2}\rightarrow 0
\ee
always corresponds to the massless limit of an infinite tower of states and to the breakdown of the effective field theory.}
\end{spacing}

\end{quote}
A natural consequence of this conjecture would be that the limit of vanishing gravitino mass, eq.~\eqref{m32limit}, is at infinite distance in the parameter space of an EFT of quantum gravity.

In the rest of the paper, we will consider a generic dependence of the mass of the tower $m$ in terms of the gravitino mass, such as 
\begin{equation}\label{masstower}
m\sim\left(m_{3/2}\right)^{n}\,,
\end{equation}
with $n$ being an order-one parameter. Note that all masses are measured in units of the $d$-dimensional effective Planck mass $M_P^{(d)}$. This dependence is found in several examples in string theory, in anti-de Sitter, Minkowski and de Sitter backgrounds. However, deviations from this simple dependence might be possible, for example logarithmic corrections as discussed in \cite{Blumenhagen:2019vgj}, but still satisfying the GMC. We argue that eq.~\eqref{masstower} should hold for $n= \mathcal{O}(1)$ and refer to the case of $n=1$ as the `strong GMC'. Evidences that the mass of an infinite tower might be related to $m_{3/2}$ have been provided also in \cite{Antoniadis:1988jn,Palti:2020tsy}.

The reason why the effective theory ceases to be valid can be understood when we identify the quantum gravity cut-off with the species scale \cite{Dvali:2007wp,Dvali:2007hz}. In fact, the cut-off of a gravitational EFT drops down in the presence of a large number of species. We will discuss this in more details in Sec.~\ref{Sec.QGcut-off}.

\subsection{The GMC and the Anti-de Sitter Distance Conjecture}\label{GMCADC}

The Anti-de Sitter Distance Conjecture (ADC) \cite{Lust:2019zwm} is obtained by comparing anti-de Sitter spaces with different cosmological constants $\Lambda$, i.e.~by varying the metric with respect to $\Lambda$. The ADC is a bit different from the SDC in the sense that it deals with the distance between non-continuously connected backgrounds, which implies that the scalar fields are not anymore massless scalars but possess a potential. Different discrete vacua are labeled by different values of the cosmological constant or the anti-de Sitter radius, which in string theory are typically related to different, discrete flux quantum numbers in the potential.

The claim of ADC is that the limit of a small AdS cosmological constant, $\Lambda\rightarrow0$,  is at infinite distance in the space of anti-de Sitter metrics, and that it is related to an infinite tower of states with typical masses that behave as
\begin{equation}
m\sim \left|\Lambda\right|^a\, ,\label{adstower}
\end{equation}
with $a={\cal O}(1)$.  The strong version of the ADC proposes that for supersymmetric anti-de Sitter vacua  $a=1/2$. 
For general anti-de Sitter backgrounds it is argued that $a\geq1/2$.

As it is well known in supergravity theories, the Lagrangian gravitino mass in a supersymmetric anti-de Sitter background is non-zero.\footnote{In this work, we will be always referring to the Lagrangian mass.}
Specifically, $m_{3/2}$ is related to the anti-de Sitter cosmological constant in the following simple way:
\begin{equation}\label{m32SUSYAdS}
m_{3/2}^2=-{\Lambda\over 3}\, .
\end{equation}

Using this relation, one can immediately see that the GMC is completely equivalent to the strong ADC for SUSY anti-de Sitter vacua, and one has that $n=2a$. For backgrounds like $AdS_d\times S^{d'}$ with  $n=1$,  there is no scale separation between $m_{3/2}$ and the tower mass scale.

Besides supersymmetric anti-de Sitter vacua, the two conjectures GMC and ADC make different predictions. Already in the case of non-supersymmetric anti-de Sitter backgrounds (despite  being arguably unstable \cite{Ooguri:2016pdq} - see also \cite{Hebecker:2020ejb} for a recent work on supersymmetry breaking in anti-de Sitter space), the relation between the gravitino mass and the cosmological constant is not one-to-one, rather there is in general a lower bound
\begin{equation}
m_{3/2}^2>{-{\Lambda\over 3}}\, .
\end{equation}
In fact, one could go to the limit of small cosmological constant $|\Lambda|\rightarrow 0$, while keeping $m_{3/2}$ finite. In this limit, the ADC predicts a tower of states with a mass scale that goes to zero following eq.~\eqref{adstower}, whereas the GMC does not, as the tower mass scale is fixed by eq.~\eqref{masstower}.

The GMC applies also to the case of Minkowski and de Sitter backgrounds, since the gravitino mass plays a central role even in such cases. On the contrary, the ADC does not apply directly to Minkowki space and its generalization to de Sitter space, still suggested in \cite{Lust:2019zwm}, makes again predictions different from the GMC.
More precisely, in the limit of small positive cosmological constant, it was argued in \cite{Lust:2019zwm} that the mass of a light tower of states would still follow the same scaling \eqref{adstower}. This leads to the strong implication that today we should expect a tower of states with extremely low mass, namely
\be\label{towerdS}
m \sim 10^{-120a}\,,
\ee
 in reduced Planck mass units. However, from the phenomenological viewpoint, it is not so easy to imagine what the origin of such a light tower of states could be;  besides in string theory it is not clear, if e.g.~KK modes
become light in the limit of small positive $\Lambda$. Moreover, the ADC predicts that the three branches, anti-de Sitter, Minkowski, and de Sitter are always separated by infinite distance in metric space and that there is no effective field theory with a finite number of fields which can have families of vacua interpolating from one branch to the other. This is a quite strong conjecture, whose validity is not yet proven.

The GMC relates instead the mass of the tower just to the mass of the gravitino, which is independent from the value of the potential when supersymmetry is broken. In the limit of small positive cosmological constant, the gravitino mass can be still large enough and we can avoid the drastic consequences of having a tower with mass scale given by eq.~\eqref{towerdS}.

\section{${\cal N}=1$ Effective Action and the GMC}\label{n1}

As we said in the last section, the GMC is argued to hold also for Minkowski or de Sitter backgrounds where supersymmetry is broken. Therefore let us consider the four-dimensional ${\cal N}=1$  effective supergravity action \cite{Cremmer:1982en}, with supersymmetry breaking F-terms and also D-terms. The low-energy degrees of freedom consist of the  ${\cal N}=1$ supergravity multiplet with the spin-2 graviton field $g_{\mu\nu}$ and the spin-3/2 gravitino $\Psi_\mu$. Then, there are $n_C$ chiral multiplets $\Phi^i=(\phi^i,\psi^i)$ ($i=1,\dots ,n_C$) and $n_V$ vector multiplets $V^a=(\lambda^a,A_\mu^a)$ ($a=1,\dots ,n_V$).
The effective action is specified by three quantities. First, the real and gauge invariant function
\begin{equation}
G(\phi,\bar\phi)=K(\phi,\bar\phi)+\log|W(\phi)|^2\, ,
\end{equation}
where $K$ is the real K\"ahler potential and $W$ the holomorphic superpotential. Second, the gauge kinetic function $f_{ab}(\phi)$. The third quantity are the Killing vectors $k_a= k_a^i(\phi)(\partial/\partial\phi^i)$ associated with the isometries of the scalar manifold that are gauged by the vector fields. They appear in the covariant derivates in the form $D_\mu \phi^i=\partial_\mu\phi^i-A_\mu^ak_a^i(\phi)$.

The scalar potential contains three terms, determined by the auxiliary fields of the gravitational, chiral and vector multiplets:
\begin{eqnarray}\label{scalarpotential}
V&=& V_G+V_F+V_D\, ,\nonumber\\
V_G    &=&   -3e^G=  -3e^K |W(\phi)|^2   \leq0         \, ,         \, \nonumber\\
V_F&=&  e^G  G^{i\bar \jmath} G_iG_{\bar \jmath} \geq0\, ,\nonumber\\
V_D  &=&   {1\over 2}  \lbrack(\Re f)^{-1}\rbrack^{ab}D_aD_b \geq0   \,.
\end{eqnarray}
Here, the $G_i$ are related to the  supergravity F-terms,
\begin{equation}
F_i\equiv e^{G/2}G_i=e^{K/2}|W|\left(K_i+\frac{W_i}{W}\right)\, ,
\end{equation}
and the gauge-invariant D-terms  are given as
\begin{equation}
D_a=iG_ik_a^i = i K_i k^i_a + \xi_a\, ,
\end{equation}
where $\xi_a$ are the so-called Fayet--Iliopoulos terms, corresponding to the gauging of the U(1) R-symmetry. For $F_i\neq 0$ or $D_a\neq 0$, ${\cal N}=1$ supersymmetry is spontaneously broken.  Note that in absence of Fayet-Iliopolous terms, the D-terms are proportional to the F-terms and  hence in this case there cannot be pure D-term breaking of supergravity, unless $|W|=0$.

The gravitino mass is given as 
\begin{equation}\label{m32def}
m_{3/2}=e^{G/2}=e^{K (\phi,\bar\phi)/2}|W(\phi)|\,,
\end{equation}
and it contributes to the scalar potential, eq.~\eqref{scalarpotential}, as
\be\label{Vm32}
V= V_F+V_D - 3 m_{3/2}^2 \,.
\ee
One sees  that
$m_{3/2}$ and $V$ are in general not anymore proportional to each other, in case supersymmetry is broken, i.e.~if either $V_F\neq 0$ or $V_D\neq 0$. In general, $m_{3/2}$ is bound by the potential from below in the following way
\begin{equation}
m_{3/2}^2\geq     -  {V\over 3}\, ,\label{vbound}
\end{equation}
where the lower limit is saturated for unbroken supersymmetry and agrees with the gravitino mass in supersymmetric anti-de Sitter backgrounds, as given by eq.~\eqref{m32SUSYAdS}. It is interesting to notice that the bound \eqref{vbound} follows from requiring unitary propagation \cite{Deser:2001pe,Zinoviev:2007ig} and holds more generally than in $\mathcal{N}=1$ supergravity. In fact, it looks like the analogous of the Higuchi bound \cite{Higuchi:1986py}, which was derived for bosonic higher spin fields following precisely unitarity considerations.

As advocated in the previous section, the GMC connects the gravitino mass (and not the value of the potential) with the mass of a tower of states. Therefore, we have to compute the relevant tower mass scale $m$ for the ${\cal N}=1$  effective supergravity action. Since supergravity is a low-energy description of string theory, it is natural to assume that the relevant tower is provided by the Kaluza-Klein (KK) particles, which arise e.g.~from a Calabi-Yau or orbifold compactification from ten to four spacetime dimensions. Then, the relevant KK mass scale, measured in units of the 4-dimensional Planck scale, for an isotropic manifold is given  as
\begin{equation}
m_{KK}=\biggl({1\over {\cal V}}\biggr)^{2/3}\,\label{kktower}
\end{equation}
where  ${\cal V}$ is the volume of the internal 6-dimensional space.\footnote{Formula \eqref{kktower} should contain also a factor $g_s^\frac14$. However, we are not keeping track of dilaton factors, as we are working in the perturbative regime of string theory by assumption.}

In the associated effective ${\cal N}=1$ supergravity theory,
 the volume ${\cal V}$  is closely related to the ${\cal N}=1$ K\"ahler potential in the following way:
\begin{equation}
K(\phi,\bar\phi)=-\alpha\log{\cal V}(\phi,\bar \phi)+K'\, .
\end{equation}
The volume ${\cal V}$ is determined by the K\"ahler moduli of the Calabi-Yau space and $\alpha$ is a model dependent parameter. For example, for heterotic string compactifications $\alpha=1$, while for type IIB orientifolds with fluxes $\alpha=2$. $K'$ is the remaining part of the K\"ahler potential, which depends on the complex structure moduli and the dilaton fields.

We also need to know how the superpotential $W$ is scaling with ${\cal V}$. For deriving this we assume that, in an extremum of the potential  with $\partial V/\partial\phi=0$, the superpotential is scaling with the volume as
\begin{equation}\label{Wscaling}
\langle W \rangle\sim{\cal V}^{\beta/2}\, ,
\end{equation}
where $\beta$ is a parameter which depends on the details of the low energy theory. The VEV symbol in eq.~\eqref{Wscaling} is to make explicit that we consider the volume-dependence of $W$ at the minimum of the potential. Note that this simple scaling behaviour of the superpotential can be modified in case of non-perturbative contributions. Neglecting the latter, we can combine our estimations for $K$ and $W$ together and we obtain for the gravitino mass the following leading behaviour in terms of the volume ${ \cal V}$:
\begin{equation}
m_{3/2}\sim\, \bigg({1\over {\cal V}}\biggr)^{{\alpha-\beta\over 2}}.\label{32vol}
\end{equation}
Comparing this with the mass scale $m_{KK}$ of the KK tower in eq.(\ref{kktower}), we can express the parameter $n$ in the GMC as:
\begin{equation}
n={4\over 3(\alpha-\beta)}\, .
\end{equation}

As said before, we argue that $n= \mathcal{O}(1)$. We will check this situation in the examples below. We will consider the case of anti-de Sitter, Minkowski and de Sitter separately.

\subsection{4D anti-de Sitter}\label{Sec.4DAdS}

As already discussed in Sec.~\ref{GMCADC}, for 4D supersymmetric  anti-de Sitter vacua, the GMC is completely equivalent to the strong ADC. One known example is M-theory on $AdS_4\times S^7$, where $n=1$.

A second class of examples are supersymmetric IIB vacua, like  the KKLT scenario \cite{Kachru:2003aw}, with a  tree-level 3-form flux superpotential plus a non-perturbative superpotential, \mbox{$W=W_0+W_{n.p.}$}, and with a K\"ahler potential $K=-3\log(T+\bar T)+\ldots =-2\log {\cal V}+ \dots$.  Hence, the isotropic Calabi-Yau volume is ${\cal V}=(\Re T)^{3/2}$.
However, the non-perturbative contribution to the superpotential provides a further exponential suppression such that $\langle W \rangle \sim T e^{-cT}$. This is reflected in the fact that a naive estimate would suggest that $m_{3/2}$ is exponentially smaller than $m_{KK}$ and the GMC, as well as the ADC, seems to be violated. Crucially, as discussed in \cite{Bena:2018fqc,Blumenhagen:2019qcg,Blumenhagen:2019vgj}, for large exponential suppression and small $W_0$ the isotropic scaling behaviour of the KK masses in eq.~\eqref{kktower} is not valid anymore, since the background becomes a highly warped throat with a different KK scaling  behaviour. Then, as pointed out in \cite{Blumenhagen:2019vgj}, also  $m_{KK}$ gets exponentially suppressed such that the supersymmetric KKLT minimum satisfies the ADC or the GMC with $a=1/6$ or with $n=1/3$ respectively, times a certain correction logarithmic in $m_{3/2}$. This is a first example where we obtain $n<1$ and the pure scaling in eq.~\eqref{masstower} is no longer valid and gets corrections.
  
In the case of non-SUSY anti-de Sitter vacua, the GMC makes in general different predictions from the ADC, as discussed in Sec.~\ref{GMCADC}.

An example of non-SUSY anti-de Sitter vacuum is provided by the Large Volume Scenario (LVS) \cite{Balasubramanian:2005zx,Conlon:2005ki}. Here, moduli stabilization is obtained by balancing $\alpha'$ corrections in the K\"ahler potential $K=-2\log(\mathcal{V} + \xi/2)$ and non-perturbative corrections in the superpotential. Up to these corrections, the superpotential is mainly given by the flux part $W_0$, which is independent of the volume and in LVS can easily be of order one (differently from KKLT). The isotropic scaling for the KK mass eq.~\eqref{kktower} is thus sensible and we have $\alpha=2$ and $\beta=0$, giving $m_{3/2}\sim\mathcal{V}^{-1}$ and $n=2/3$. Also in this case, the non-perturbative contributions in the superpotential leads to logarithmic corrections in the dependence in terms of the cosmological constant \cite{Blumenhagen:2019vgj}.

The distribution of the gravitino mass in KKLT and LVS has been recently investigated in \cite{Broeckel:2020fdz}. The results of this work provide further support for our conjecture.

\subsection{4D Minkowski}

Let us now discuss a few cases of 4D Minkowksi vacua with spontaneously broken ${\cal N}=1$ supersymmetry. 
Since the potential is zero\footnote{Of course all known string constructions with broken supersymmetry are plagued by the fact that a non-zero potential is created by string loop effects, as e.g.~discussed already in \cite{Rohm:1983aq}. So in our paper we do not claim that non-supersymmetric (no-scale) string models with all order zero cosmological constant easily exist, they might still belong to the swampland.}, 
the ADC cannot be applied. However, due to the breaking of supersymmetry, the gravitino mass is non-vanishing and the GMC still predicts a tower of light states in the limit of $m_{3/2}\rightarrow 0$. In the case of Minkowki space, the GMC predicts the breakdown of the effective theory in the limit of small supersymmetry breaking scale.

\subsubsection{F-term no-scale models}

The simplest  4D no-scale models \cite{Cremmer:1983bf,Ellis:1983ei,Ellis:1984bm}
with spontaneously broken supersymmetry and vanishing potential are given by a K\"ahler potential of the  form
\begin{equation}
K=-3\log(T+\bar T)+ K'(\phi,\bar\phi)
\end{equation} 
and a superpotential that does not depend on the $T$, e.g.~$W={\rm const}$. Supersymmetry is spontaneously broken by the F-ferm $F_T$. For heterotic string compactifications the volume is given as ${\cal V}=(T+\bar T)^3$ and thus we find that $\alpha=1$ and $\beta=0$. It follows that the GMC is satisfied with $n=4/3$. For type IIB GKP orientifolds \cite{Giddings:2001yu} we have that $\alpha=2$, $\beta=0$ and hence $n=2/3$.

\subsubsection{Scherk-Schwarz compactifications}

Another class of 4D Minkowski vacua with broken supersymmetry is given by coordinate dependent string Scherk-Schwarz compactifications \cite{Scherk:1978ta,Scherk:1979zr}. They are also of the no-scale type and can be realized in string theory, with supersymmetry is broken at tree level. Several examples of this type were  
studied in \cite{Ferrara:1994kg}. They are characterized by a K\"ahler potential
\begin{equation}
K=-\log\lbrack(S+\bar S)(T+\bar T)(U+\bar U)\rbrack+ K'(\phi,\bar\phi)\, ,
\end{equation} 
where $S$ is the complexified heterotic dilaton, while $T$ and $U$ are the K\"ahler and complex structure moduli of a two-torus. The effective Scherk-Schwarz superpotential  is constant at string tree level and leads to three non-vanishing F-terms, $F_S$, $F_T$ and $F_U$. This choice of K\"ahler and superpotential corresponds to $\alpha=1/3$ and $\beta=0$ leading to $n=4$. These  so called STU models possesses an underlying extended ${\cal N}=2$ supersymmetry structure, which will be further discussed in section \ref{n2}.

 \subsubsection{F-term and D-term supersymmetry breaking}
 \label{DZ}

Another interesting class of Minkowski models, where supersymmetry is broken by an F-term and a D-term and with a no-scale potential was introduced in \cite{DallAgata:2013jtw}. Here one has a K\"ahler potential of the form
\begin{equation}
K=-2\log(T+\bar T)+ K'(\phi,\bar\phi)
\end{equation} 
which corresponds to $\alpha=2/3$ and again constant superpotential $W_0$. In addition there is an Abelian vector multiplet with gauge coupling constant $f={1\over g^2}$ that gauges the shift symmetry associated to the axion $i(T-\bar T)$. The corresponding holomorphic Killing vector is an imaginary constant, $k=iq$. Then, one can show \cite{DallAgata:2013jtw}  that the total  scalar potential $V= V_G+V_F+V_D$ is everywhere vanishing provided one identifies
\begin{equation}
|W_0|=\sqrt 2 g|q|\, .\label{iden}
\end{equation}
As a consequence, the gravitino mass is proportional to the gauge coupling constant $g$,
\begin{equation}
\label{m=g}
m_{3/2}={gq\over \sqrt 2\Re T}\, ,
\end{equation}
and the GMC is satisfied with $n=2$. Interestingly, in the limit of vanishing gauge coupling, $g\rightarrow0$, also $m_{3/2}$ is vanishing. In this case, one can thus interpret the GMC as an obstruction in the restoration of a global symmetry, something which is believed to be incompatible with quantum gravity.  The relation between $m_{3/2}$ and the gauge coupling will be further explored in section  \ref{n2}.

\subsection{4D de Sitter}

Next let us consider the possibility of having de Sitter vacua in quantum gravity or in string theory, opposite to the various arguments against de Sitter vacua quoted in the introduction of our paper.

Given the form of the scalar potential eq.~\eqref{Vm32} in a 4D $\mathcal{N}=1$ EFT, we see that the gravitino mass sets the relation between the scale of supersymmetry breaking and the value of the cosmological constant. The limit of small positive cosmological constant can therefore correspond to a finite value of $m_{3/2}$, if this is properly compensated by a finite amount of supersymmetry breaking. Then, in this limit the GMC does not predict any light tower, differently from the ADC which relates the mass of the tower to value of the cosmological constant itself (see section \ref{GMCADC}).

Although we have conjectured that the GMC should hold independently from the specific background, we present below a simple argument to show that the applicability of the GMC to the case of de Sitter can also be supported by its validity in anti-de Sitter spaces (in the case of supersymmetric anti-de Sitter, it directly corresponds to the validity of the ADC). In fact, some of the best and most studied de Sitter constructions from string theory \cite{Kachru:2003aw,Balasubramanian:2005zx,Westphal:2006tn} make use of specific objects, such as anti-D-branes, to uplift an anti-de Sitter vacuum with potential value $V_{AdS}$ and yield a positive minimum, with potential\footnote{The use of anti-branes, as first considered for uplift, inflation
 and brane supersymmetry breaking in \cite{Dvali:1998pa,Antoniadis:1999xk,Dvali:2001fw},
has been under severe scrutiny during the last years. On the one hand, considerable progress has been made in the understanding of the correct description at the EFT level in terms of non-linear supersymmetry with constrained superfields \cite{Ferrara:2014kva,Kallosh:2014wsa,Polchinski:2015bea,Vercnocke:2016fbt,McDonough:2016der,GarciadelMoral:2017vnz,Kallosh:2018wme,Cribiori:2019hod,Parameswaran:2020ukp,Cribiori:2020bgt}. On the other hand, possible issues have been identified when using warped throats \cite{Bena:2018fqc,Blumenhagen:2019qcg,Scalisi:2020jal}, which are useful to create energy hierarchies and redshift the tension of the anti-branes.}
\be
V_{dS}=V_{AdS} + V_{up}\,.
\ee
The peculiarity of such uplifting terms is that their contribution scales as a negative power of the internal volume, such as
\be\label{Vup}
V_{up}=\frac{c}{\mathcal{V}^p}\,,
\ee
with $c$ and $p$ positive. Assuming this dependence, one can show that the position $\mathcal{V}_1$ of the minimum of $V_{dS}$ remains almost unchanged with respect to the position $\mathcal{V}_0$ of the anti-de Sitter vacuum (see Appendix \ref{AppendixUPdS} for the derivation). As a consequence, the gravitino mass \mbox{$m_{3/2}(\mathcal{V})= e^{K(\mathcal{V})/2}|W(\mathcal{V})|$} will not change considerably after uplifting and one will have
\be
m_{3/2}(\mathcal{V}_1)\simeq m_{3/2}(\mathcal{V}_0)\,.
\ee

Therefore, in the situation where the de Sitter vacuum is obtained from anti-de Sitter via an uplifting of the form \eqref{Vup}, the consequences of the GMC in de Sitter directly follow from the discussion of the GMC in anti-de Sitter.

Primary examples of this situation are the KKLT and LVS de Sitter vacua. Both of these constructions have an infinite KK tower with mass scale going to zero in the limit of small gravitino mass $m_{3/2}\rightarrow 0$ and they therefore satisfy the GMC.  In fact, the same analysis presented in Sec.~\ref{Sec.4DAdS} can be applied also to the de Sitter case, after uplifting.

\section{${\cal N}=2$ Effective Action and the GMC}\label{n2}

In $\mathcal{N}=1$ supergravity the gravitino mass is in general independent from the gauge kinetic function. The situation is different in theories with more supersymmetry. Our motivation to discuss now $\mathcal{N}=2$ supergravity is indeed that the gravitino mass is necessarily related to the gauge coupling and this will allow us to connect the GMC to other swampland conjectures.

\subsection{Review of $\mathcal{N}=2$ supergravity}

We consider 4D $\mathcal{N}=2$ supergravity, mostly following the conventions of \cite{Andrianopoli:1996cm}. The low-energy degrees of freedom consist of the $\mathcal{N}=2$ supergravity multiplet with the spin-2 graviton field $g_{\mu\nu}$, two spin-3/2 gravitini $\Psi_{\mu}^A$ and the spin-1 graviphoton $A^0_\mu$. Then, there are $n_V$ vector multiplets $(z^i, \lambda_A^i, A_\mu^i)$ and $n_H$ hypermultiplets $(q^u, \zeta^\alpha)$.

The scalars $z^i$ in the vector multiplets are coordinates of a special K\"ahler manifold, while the scalars $q^u$ of the hypermultiplets are coordinates of a quaternionic manifold, which in general is not K\"ahler. The total scalar manifold is the product of these two spaces.

\subsubsection{Special K\"ahler geometry, killing vectors and prepotentials}

The vector multiplets scalars are coordinates of a special K\"ahler manifold, with metric $g_{i \bar \jmath}$. The manifold is endowed with a U(1) Lie-algebra valued 2-form which is closed, namely the K\"ahler form. 

A convenient formulation of Special K\"ahler geometry is given in terms of the holomorphic sections $(X^\Lambda (z), F_\Lambda (z))$, $\Lambda=0,1,\dots,n_V$, entering the K\"ahler potential
\begin{equation}
K = - \log [i (\bar X^\Lambda F_\Lambda - \bar F_\Lambda X^\Lambda)].
\end{equation} 
One can also define covariantly-holomorphic sections $(L^\Lambda, M_\Lambda)=e^\frac{K}{2}(X^\Lambda, F_\Lambda)$ and their derivatives $f_i^\Lambda = D_i L^\Lambda$, $h_{\Lambda\, i}  = D_i M_\Lambda$. The covariant derivatives are defined as $D_i X^\Lambda = (\partial_i+K_i )X^\Lambda$ and $D_i L^\Lambda = (\partial_i + \frac12 K_i ) L^\Lambda$. These quantities satisfy several interesting properties, for which we refer to \cite{Ceresole:1995ca}. Here, we recall the pullback of the inverse K\"ahler metric
\begin{equation}
U^{\Lambda \Sigma} \equiv f_i^\Lambda g^{i \bar \jmath} f_{\bar \jmath}^\Sigma = -\frac12 ({\rm Im }\,\mathcal{N}^{-1})^{\Lambda \Sigma} - \bar L^{\Lambda} L^\Sigma,
\end{equation}
where the complex symmetric period matrix $\mathcal{N}_{\Lambda \Sigma}$ is defined such that $M_\Lambda = \mathcal{N}_{\Lambda \Sigma} L^\Sigma$. It enters the kinetic terms of the vector fields
\begin{equation}
e^{-1}\mathcal{L}_{kin} = \frac14 {\rm Im}\,\mathcal{N}_{\Lambda \Sigma}\,\,F^{\Lambda}_{\mu\nu} F^{\Sigma\,\mu\nu} + \frac18 {\rm Re}\,\mathcal{N}_{\Lambda \Sigma}\,\,\epsilon^{\mu\nu\rho\sigma} F_{\mu\nu}^{\Lambda}F_{\rho \sigma}^{\Sigma}.
\end{equation}

Isometries are generated by holomorphic killing vectors $k^i_\Lambda (z)$, satisfying the algebra
\begin{equation}
\label{gaugealgebra}
[k_\Lambda, k_\Sigma] = - f^{\Delta}_{\Lambda \Sigma} k_\Delta.
\end{equation}
Explicitly, they are given in terms of a real prepotential function $\mathcal{P}^0_\Lambda$ as
\begin{equation}
k^i_\Lambda = i g^{i \bar \jmath} \partial_{\bar \jmath} \mathcal{P}^0_\Lambda,
\end{equation}
and the prepotential is such that $\mathcal{P}^0_\Lambda L^\Lambda =0$. For consistency, they have to satisfy the equivariance condition
\begin{equation}
i g_{i \bar \jmath} (k^i_\Lambda k^{\bar \jmath}_\Sigma - k^i_\Sigma k^{\bar \jmath}_\Lambda) = - f^{\Delta}_{\Lambda \Sigma} \mathcal{P}^0_\Delta.
\end{equation}

\subsubsection{Quaternionic geometry, killing vectors and prepotentials}

The hypermultiplets scalars $q^u$ are coordinates of a quaternionic manifold with metric $h_{uv}$. The manifold is endowed with a  SU(2)-Lie algebra valued 2-form $\Omega^x$, $x=1,2,3$, which is not closed but covariantly closed
\begin{equation}
\nabla \Omega^x \equiv d \Omega^x + \epsilon^{xyz} \omega^y \wedge \Omega^z =0,
\end{equation}
where $\omega^x$ is a connection, namely $\Omega^x = d \omega^x + \frac12 \epsilon^{xyz} \omega^y \wedge \omega^z.$ The non-closure of $\Omega^x$ reflects the fact that the quaternionic manifold is not K\"ahler. The components of $\Omega^x$ provide a representation of the quaternionic algebra, i.e.
\begin{equation}
h^{st}\Omega^x_{us}\Omega^y_{tw} = - \delta^{xy}h_{uw}-\epsilon^{xyz}\Omega^z_{uw}.
\end{equation}

Isometries are generated by killing vectors $k^\mu_{\Lambda}(q)$ satisfying the same algebra as in \eqref{gaugealgebra}. They can be given in terms of a triplet of real prepotentials $\mathcal{P}^x_\Lambda$ as 
\begin{equation}
k^u_\Lambda = \frac16 \Omega^{x, uv} \nabla_v \mathcal{P}^x_\Lambda
\end{equation}
and are solution of $2 \Omega_{uv} k^u_\Lambda = (\partial_v \mathcal{P}^x_\Lambda + \epsilon^{xyz}\omega^y_v \mathcal{P}^z_\Lambda)\equiv\nabla_v \mathcal{P}^x_\Lambda$. Explicitly, the prepotentials can be expressed in terms of the connection $\omega^x$ and of a SU(2)-compensator $C^x_\Lambda$ as \cite{DAuria:1990qxt}
\begin{equation}
\mathcal{P}^x_\Lambda = C^x_\Lambda - k^u_\Lambda \omega_u^x.
\end{equation}
The equivariance consistency condition for the quaterionic manifold is
\begin{equation}
2 k^u_\Lambda k^v_\Sigma \Omega^x_{uv} = - f^\Gamma_{\Lambda \Sigma} \mathcal{P}^x_\Gamma - \epsilon^{xyz} \mathcal{P}^y_\Lambda \mathcal{P}^z_\Sigma.
\end{equation}

\subsection{The scalar potential, the gravitino gauge coupling and mass}

The only way to generate a scalar potential in extended supergravity is by means of a gauging. In the case of $\mathcal{N}=2$ supergravity, the scalar potential arising from a generic (electric) gauging has the form
\begin{equation}
\label{VN=2}
{V}_{\mathcal{N}=2} = {V}_1+ {V}_2+{V}_3,
\end{equation}
where
\begin{align}
{V}_1 &= g_{i \bar \jmath} k^i_\Lambda k^{\bar \jmath}_\Sigma \bar L^\Lambda L^\Sigma,\\
{V}_2 &=4 h_{uv}k^u_\Lambda k^v_\Sigma \bar L^\Lambda L^\Sigma,\\
{V}_3 &=  (U^{\Lambda \Sigma} - 3\bar L^\Lambda L^\Sigma) \mathcal{P}_\Lambda^x \mathcal{P}_\Sigma^x.
\end{align}
The first and second terms arise from the gauging of isometries on the Special K\"ahler and quaternionic manifold respectively. The third term contains the contribution from Fayet--Ilioupoulos terms.

It might be instructive to rewrite the scalar potential in a language which is reminiscent of $\mathcal{N}=1$ supergravity, namely in terms of a superpotential $\mathcal{W}$ and D-terms $\mathcal{D}^\Lambda$. To identify the superpotential, we can look at the gravitino mass matrix
\begin{equation}
\label{SAB}
S_{AB} = \frac i2 \, {(\sigma_x)_A}^C \epsilon_{BC} \mathcal{P}^x_\Lambda L^\Lambda = -\frac i2 \left(\begin{array}{cc}
 \mathcal{P}_\Lambda^1 L^\Lambda-i  \mathcal{P}_\Lambda^2 L^\Lambda & - \mathcal{P}_\Lambda^3 L^\Lambda\\
 - \mathcal{P}_\Lambda^3 L^\Lambda &  -\mathcal{P}_\Lambda^1 L^\Lambda - i  \mathcal{P}_\Lambda^2 L^\Lambda\\
 \end{array}
 \right).
\end{equation}
One natural possibility is to identify the superpotential as one of the diagonal entries. Since we have the freedom of choosing an SU(2) orientation, for definiteness we set \cite{Andrianopoli:2001zh,Andrianopoli:2001gm, Lust:2005bd}
 \begin{equation}
 \mathcal{W}(z,q) = \mathcal{P}_\Lambda (q)L^\Lambda(z), \qquad W(z,q)  =   \mathcal{P}_\Lambda (q)X^\Lambda(z), \qquad \mathcal{P}_\Lambda \equiv \mathcal{P}_\Lambda^1 - i \mathcal{P}_\Lambda^2.
 \end{equation}
 As for the D-terms, the proper identification is then
  \begin{equation}
 \label{Dtermdef}
 \mathcal{D}^\Lambda (z,q) = ({\rm Im}\, \mathcal{N}^{-1})^{\Lambda \Sigma} (\mathcal{P}^0_\Sigma (z)+ \mathcal{P}^3_\Sigma (q)).
 \end{equation}
By exploiting several properties of special K\"ahler and quaternionic geometry, one can then rewrite the complete $\mathcal{N}=2$ scalar potential as
\begin{equation}
\label{V2final}
{V}_{\mathcal{N}=2} = {V}_{\mathcal{N}=1} -2\bar L^\Lambda L^\Sigma(h_{uv}k^{u}_\Lambda k^v_\Sigma + 2 \mathcal{P}_\Lambda^3 \mathcal{P}_\Sigma^3),
\end{equation}
where we defined a seemingly $\mathcal{N}=1$ scalar potential
\begin{equation}
{V}_{\mathcal{N}=1} = {V}_F + V_G+{V}_D
\end{equation}
with contributions analogous to the ones in section \ref{n1}:
\begin{align}
{V}_{F}& =  e^K \left(g^{i  \bar \jmath} D_i {W} D_{\bar \jmath} \bar{{W}}  + \frac12 h^{uv} \nabla_u \bar{{W}}\nabla_v {W} +\frac12 h^{uv}  \nabla_u (\bar X^\Lambda \mathcal{P}^3_\Lambda ) \nabla_v (X^\Sigma \mathcal{P}^3_\Sigma)\right), \\
V_G&=- 3 e^K \bar{{W}} {W},\\
\label{VD2}
{V}_D &= -\frac12{\rm Im} \mathcal{N}_{\Lambda \Sigma}\, \mathcal{D}^\Lambda \mathcal{D}^\Sigma.
\end{align}
The minus sign in $V_D$ is consistent with the fact that the matrix ${\rm Im} \mathcal{N}_{\Lambda \Sigma}$ is negative definite. More details on the calculation can be found in the Appendix \ref{appB}. We stress that this is just a formal rewriting, since a fully-fledged description of $\mathcal{N}=2$ supergravity in terms of $\mathcal{N}=1$ quantities would require necessarily a restriction on the quaternionic manifold to become K\"ahler \cite{Andrianopoli:2001zh,Andrianopoli:2001gm}.

Another effect of the gauging is that the gravitini are charged. Their gauge coupling and charge can be read off from the covariant derivative
\begin{equation}  
D_\mu \Psi_{\nu\, A} = \dots +\frac i2  A_\mu^\Lambda \mathcal{P}^0_\Lambda \Psi_{\nu\, A} + \frac i2  A_\mu^\Lambda \mathcal{P}^x_\Lambda {(\sigma_x)_A}^B \Psi_{\nu \, B},
\end{equation}
once the vector fields have been canonically normalized. For simplicity, we restrict only to a U(1)$_R$ gauging by means of the hypermultiplet prepotentials, thus we set $\mathcal{P}^0_\Lambda =0$. Without loss of generality, we can rotate the SU(2) frame such that
\begin{equation}
\mathcal{P}^x_\Lambda = e_\Lambda \delta^{1x},
\end{equation}
where $e_\Lambda$ can depend on the scalars $q^u$. The vector field responsible for the gauging is then given by the combination
\begin{equation}
\tilde A_\mu = \Theta_\Lambda A^\Lambda_\mu, \qquad \Theta_\Lambda \equiv \frac{e_\Lambda}{2q_{3/2}},
\end{equation}
where $\pm q_{3/2}$ is the charge of the gravitini, which we assume to be non-vanishing (and quantized). To correctly identify the gauge coupling we have to canonically normalise the vector fields. To this purpose, we introduce the projectors \cite{Cribiori:2020use}
\begin{equation}
{{P^\parallel}^\Lambda}_\Sigma= \frac{({\rm Im} \mathcal{N}^{-1})^{\Lambda \Gamma}\Theta_\Gamma \Theta_\Sigma}{\Theta^2}, \qquad {{P^\perp}^\Lambda}_\Sigma = \delta^{\Lambda}_\Sigma - {{P^\parallel}^\Lambda}_\Sigma,
\end{equation}
where $\Theta^2 \equiv \Theta_\Lambda ({\rm Im} \mathcal{N}^{-1})^{\Lambda \Sigma}\Theta_\Sigma $, and we split the vector fields as
\begin{equation}
A_\mu^\Lambda = A^{\perp\, \Lambda}_\mu + \frac{({\rm Im} \mathcal{N}^{-1})^{\Lambda \Sigma}\Theta_\Sigma}{\Theta^2} \tilde A_\mu, \qquad A^{\perp\, \Lambda}_\mu \equiv {{P^\perp}^\Lambda}_\Sigma A^\Sigma_\mu.
\end{equation}
Substituting this expression into the kinetic term, we get then
\begin{equation}
\frac14 ({\rm Im} \mathcal{N})_{\Lambda \Sigma} F^\Lambda_{\mu\nu} F^{\Sigma\, \mu\nu} = \frac14 ({\rm Im} \mathcal{N})_{\Lambda \Sigma} F^\Lambda_{\mu\nu} (A^\perp) F^{\Sigma\, \mu\nu} (A^\perp) + \frac14 \frac{1}{\Theta^2} F_{\mu\nu} (\tilde A) F^{\mu\nu} (\tilde A),
\end{equation}
from which we identify the gauge coupling of the gravitini
\begin{equation}
\label{g32U1}
g_{3/2} = \sqrt{-\Theta^2} = \sqrt{- \Theta_\Lambda ({\rm Im} \mathcal{N}^{-1})^{\Lambda \Sigma}\Theta_\Sigma}\,\, .
\end{equation} 
In the case in which the abelian gauging is performed with the special K\"ahler prepotential, one obtains again an expression analogous to \eqref{g32U1}, with $e_\Lambda$ replaced by $\mathcal{P}^0_\Lambda$.
For non-constant prepotentials $\mathcal{P}^x_\Lambda$, the projector $\Theta_\Lambda$ is a function of the quaternionic scalars $\Theta_\Lambda = \Theta_\Lambda (q)$, while the matrix $\mathcal{N}_{\Lambda \Sigma}$ depends on the vector multiplets scalars
\begin{equation}
-({\rm Im} \mathcal{N}^{-1})^{\Lambda \Sigma} = 2f_i^\Lambda g^{i \bar \jmath}f_{\bar\jmath}^\Sigma +2 \bar L^\Lambda L^\Sigma = 2e^K \left(D_i X^\Lambda g^{i\bar \jmath} D_{\bar \jmath} \bar X^{\Sigma} + \bar X^\Lambda X^\Sigma\right).
\end{equation}
Thus, we see that the gauge coupling of the gravitini has the general form
\begin{equation}
\label{g32gen}
g_{3/2} = e^{\frac K2} \mathcal{F}(z,q), \qquad \mathcal{F}(z,q) = \sqrt{2 \Theta_\Lambda \left(D_i X^\Lambda g^{i\bar \jmath} D_{\bar \jmath} \bar X^{\Sigma} + \bar X^\Lambda X^\Sigma\right) \Theta_\Sigma}
\end{equation}
where $\mathcal{F}(z,q)$ is a model dependent function of the scalar fields $z,q$. Crucially, there is an overall factor $e^{\frac K2} \sim \mathcal{V}^{-\frac{\alpha}{2}}$, indicating that the limit $g_{3/2} \to 0$ leads to decompactification (assuming the function $\mathcal{F}(z,q)$ is well behaved) and thus the restoration of the U$(1)_R$ global symmetry is obstructed. The relation $g_{3/2} \sim e^\frac{K}{2}$ is also suggestive from an $\mathcal{N}=1$ perspective. Indeed, the part of the $\mathcal{N}=1$ scalar potential which is not generated by a gauging has nevertheless an overall factor $e^K \sim g_{3/2}^2$. Our analysis suggests that this might be interpreted as a remnant of the gauging in a parent $\mathcal{N}=2$ theory. We hope to investigate this aspect further in the future.

Finally, notice that the gravitino mass matrix \eqref{SAB} contains a factor $e^{\frac{K}{2}}$. We can replace it with the expression \eqref{g32gen} to obtain a direct connection between $S_{AB}$ and $g_{3/2}$. The precise relation is in general model dependent, but we can look once more at specific examples. In the simple case in which the U$(1)_R$ gauging is performed with the graviphoton only, we have
\begin{equation}
-({\rm Im} \mathcal{N}^{-1})^{00} =2 e^K (K_i K_{\bar \jmath} g^{i \bar \jmath} + 1)
\end{equation}
and thus we get (normalizing $\Theta_0=1/\sqrt{2}$)
\begin{equation}
g_{3/2} = e^{\frac K2} \sqrt{K_i K_{\bar \jmath} g^{i \bar \jmath} + 1}, \qquad S_{AB} = \frac i2 {(\sigma_1)_A}^C \epsilon_{BC} \frac{q_{3/2}\, g_{3/2}}{\sqrt{K_i K_{\bar \jmath} g^{i \bar \jmath} + 1}} \,.
\end{equation}
In no-scale models these formulae simplify further, since $K_i K_{\bar \jmath} g^{i \bar \jmath}=3$. We see that a non-vanishing gravitino mass prevents us from taking the limit $g_{3/2} \to 0$ and thus from restoring a global symmetry (assuming charge quantization). This examples relates thus explicitly the GMC to the absence of global symmetries in quantum gravity. 

Recently, it was observed in \cite{Cribiori:2020use} that de Sitter vacua in $\mathcal{N}=2$ supergravity with a vanishing gravitino mass are in contradiction with the magnetic WGC and thus in the swampland.\footnote{The same argument can be used to show that pure Fayet--Ilioupoulos terms in $\mathcal{N}=1$ supergravity are in the swampland \cite{Cribiori:2020wch}.} In particular, restricting to the case with only vector multiplets, in \cite{Cribiori:2020use} a general proof is given for abelian gauging, while all known examples of non-abelian gaugings are analysed and a new class is constructed. All of them are found to clash with the magnetic WGC and to have $\det S_{AB}=0$, which is proposed as a swampland criterion for de Sitter vacua. In this respect, the GMC can thus provide a rationale behind this a priori unexpected correlation between a vanishing gravitino mass, the WGC and the conjectured absence of de Sitter vacua.

\subsection{STU model}

Since the previous discussion in $\mathcal{N}=2$ supergravity has been fairly general, it is illustrative to understand how those features arise in an explicit construction. As an example, we consider a simple model with $n_V=3$ vector multiplets and prepotential
\begin{equation}
F = \frac{X^1 X^2 X^3}{X^0}.
\end{equation}
We choose the normal coordinates
\begin{equation}
z^i = \frac{X^i}{X^0},  \quad i=1,2,3, \qquad \text{with} \quad X^0 \equiv 1.
\end{equation}
The K\"ahler potential is given by
\begin{equation}
K = - \log (s t  u)\, ,
\end{equation}
where $s = -2 {\rm Im} z^1$, $t = -2 {\rm Im} z^2$, $u= -2 {\rm Im} z^3$. In the following we set the real parts of $z^i$ to zero for convenience, ${\rm Re} z^i \equiv 0$. 
This is the so called STU model.\footnote{A concrete $\mathcal{N}=2$ example of an STU model is heterotic string compactification on $T^2\times K_3$. As discussed in
\cite{deWit:1995dmj},
the fields $t$ and $u$ correspond to the K\"ahler modulus and the complex structure modulus of $T^2$, respectively, and $s$ denotes the complexified heterotic dilaton. The $U(1)$ gauge group, discussed here,
originates from one of the four $U(1)$'s, related to the compacification on $T^2$. Moreover, the torus compactification leads to a tower of states, which is linked to the gravitino mass and to the $U(1)$ gauge coupling in the way we describe in our paper.}

We want to gauge a U(1)$_R \subset$ SU(2)$_R$ by means of the graviphoton. For definiteness, we choose 
\begin{equation}
\mathcal{P}_\Lambda^x = q_{3/2} \delta^{1x}\eta_{0\Lambda}, \qquad \text{such that} \qquad 2\Theta_\Lambda = \eta_{0\Lambda}.
\end{equation}
With this choice, we have
\begin{equation}
e^{\frac K2} = \frac{1}{\sqrt{stu}}, \qquad W=X^\Lambda \mathcal{P}_\Lambda = q_{3/2}, \qquad \mathcal{D}^\Lambda =0.
\end{equation}
One can easily check that this gauging gives rise to a no-scale scalar potential
\begin{equation}
V = 0,
\end{equation}
therefore the fields $s$, $t$, $u$ are flat directions and we have an infinite family of (non-supersymmetric) Minkowski vacua. The gauge kinetic matrix of the vectors reads
\begin{equation}
{\rm Im} \mathcal{N}_{\Lambda \Sigma} = - s t u\,\, {\rm diag}\left(\frac18,\frac{1}{2s^2},\frac{1}{2t^2},\frac{1}{2u^2}\right),
\end{equation}
while gauge coupling of the gravitini is
\begin{equation}
g_{3/2} = \sqrt{-\Theta^2} = \sqrt{\frac{2}{stu}}
\end{equation}
and the gravitino mass matrix is given by
\begin{equation}
\label{S=g}
S_{AB} = \frac i2 {(\sigma_1)_A}^C \epsilon_{BC} e^{\frac K2} W = \frac i2 \frac{q_{3/2}}{\sqrt{s  t  u}} \text{diag}(1,-1) = \frac{i}{2 \sqrt 2} q_{3/2} g_{3/2} \text{diag}(1,-1).
\end{equation}
Assuming the gravitino charge to be quantized, we see that the limit of vanishing gravitino mass implies a vanishing gauge coupling:
\begin{equation}
m_{3/2} \to 0 \qquad \Leftrightarrow \qquad g_{3/2} \to 0.
\end{equation}
This is unacceptable from a swampland perspective, since it leads to the restoration of a global symmetry. Thus, we conclude that in this model the GMC is a direct consequence of the absence of global symmetries in quantum gravity. Alternatively, one can interpret the same fact by means of the WGC, since \eqref{S=g} identifies the gravitino mass with the gauge coupling, up to constant factors. Notice that the same feature appears also in the $\mathcal{N}=1$ example analysed in section (\ref{DZ}), see in particular eq.~\eqref{m=g}.

\section{Gravitino and Quantum Gravity Cut-off}\label{Sec.QGcut-off}

We want now to use the GMC to extract information about the quantum gravity cut-off in a given effective theory. This will be relevant for the discussion on the phenomenological implications of the GMC. For convenience, in what follows we reinstate explicitly in all formulae the reduced Planck mass $M_P$, which has been set to unity so far.

Denoting with $N$ the number of light particles (or species) in a given theory and assuming the GMC to hold, in the limit of small gravitino mass, eq.~\eqref{m32limit}, $N$ will increase leading to a reduction of the quantum gravity cut-off such as
\be
\Lambda_{QG}=\frac{M_P}{\sqrt{N}}\,.
\ee
This is in fact the so-called `species scale' \cite{Dvali:2007wp,Dvali:2007hz}, above which quantum gravitational effects cannot be neglected and gravity becomes strongly coupled.

If we assume that the states of the tower are equally spaced (as it is the case for KK or winding modes), below the cut-off their number will be
\be
N=\frac{\Lambda_{QG}}{m}\,.
\ee

In this work, we have argued that a generic simple scaling of the mass of the tower $m$ in terms of $m_{3/2}$ is given by eq.~\eqref{masstower}. Once we reinstate $M_P$ explicitly, this relation reads
\be\label{massm32Mp}
m\sim M_P\left(\frac{m_{3/2}}{M_P}\right)^n\,.
\ee

Putting these last three equations together, we can derive
\be\label{QGcutoff}
\Lambda_{QG}\simeq M_P\ \left(\frac{m_{3/2}}{M_P}\right)^{\frac{n}{3}}\,,
\ee
which sets how the quantum gravity cut-off depends on the mass of the gravitino. It is also possible to write this mass in terms of the number of light species, such as
\be
m_{3/2} \simeq \frac{M_P}{N^{\frac{3}{2n}}}\,,
\ee
which is telling us that the mass of the gravitino is below the Planck scale any time a number of light particles have mass below the cut-off. In (quasi-)Minkowski space, the last equation is therefore suggesting that the scale of supersymmetry breaking should be below the Planck scale.  A detection of the gravitino would not just give us information about the scale above which weakly coupled Einstein gravity breaks down, but also about the number of light states, which could potentially produce observational effects in our EFT. 

From the latter expressions, we notice that $m_{3/2}$ will be always lower than the quantum gravity cut-off $\Lambda_{QG}$ if $n<3$. For $n=1$, we have for example
\be\label{m32N}
m_{3/2}\simeq\frac{M_P}{N^{\frac{3}{2}}}= \frac{\Lambda_{QG}^3}{M_P^2}<\Lambda_{QG}\,.
\ee

The gravitino mass will be of the order of the cut-off just for $n=3$. This means that for $n\geq 3$ we do not have any effective theory of supersymmetry breaking. \footnote{Note that the work \cite{Palti:2020tsy} claims that there is no EFT of supersymmetry breaking  when $n=1$. However, this case corresponds to a tower mass scale $m\sim m_{3/2}$, which is generically lower than the cut-off $\Lambda_{QG}$.}

In the case of a charged gravitino with gauge coupling $g_{3/2}$, we have the bound
\be
\Lambda_{QG}< g_{3/2}\ M_P\,,
\ee
imposed by the \textit{magnetic WGC}. Using the expression for the cut-off eq.~\eqref{QGcutoff}, we can derive the following relation between the mass and the gauge coupling
\be\label{WGCbound}
m_{3/2} < (g_{3/2})^{3/n} M_P < g_{3/2}\ M_P\,.
\ee
The last inequality is true if $n<3$ (assuming $g_{3/2}<1$) and it is precisely of the form suggested by the \textit{electric WGC}. It is intriguing to notice that the domain for which an effective theory of supersymmetry breaking is still consistent (as we explained after eq.~\eqref{m32N}) assures validity of the WGC for the gravitino.

\section{Gravitino and (Quasi-)de Sitter Space}\label{sec6}

In this section, we will consider the phenomenological implications of the GMC in a background with a positive cosmological constant (CC). We will apply these results mainly to the case of {\it inflation} (i.e.~with a slowly varying CC), where we might appreciate the phenomenological power of such statements. However, the results will hold more generally and they could apply also to the case of dS vacua or quintessence.

The mass of the gravitino plays an important role also in a phase with a positive potential. As discussed in sec.~\ref{n1}, in a $\mathcal{N}=1$  $D=4$ EFT, it provides the negative contribution to the scalar potential that adds to the positive SUSY breaking terms. However, given this structure (see eq.~\eqref{Vm32}), the mass of the gravitino does not constrain the value of the scalar potential itself.\footnote{Constraints between $m_{3/2}$  and the value of a positive scalar potential have been found in literature when considering other mechanisms, such as destabilization of the volume modulus \cite{Kallosh:2004yh,Conlon:2008cj} or catastrophic gravitinos production \cite{Hasegawa:2017hgd,Hasegawa:2017nks} (see also \cite{Kolb:2021nob,Kolb:2021xfn,Dudas:2021njv}).} A higher SUSY breaking scale might compensate a bigger gravitino mass in order to give the same result. Once we fix a value of the scalar potential, $m_{3/2}$ sets in fact the scale at which supersymmetry is broken.

A requirement for perturbative control of our EFT of de Sitter, with Hubble parameter $H$, is that the cut-off satisfies
\be
\Lambda_{QG}>H\,.
\ee
In the previous section, we have shown that a direct result of the GMC is that the quantum gravity  cut-off is set by the gravitino mass scale following eq.~\eqref{QGcutoff}. This implies a bound on the mass of the gravitino in terms of the Hubble parameter such as
\be\label{boundHm32}
m_{3/2}> M_P^{\frac{n-3}{n}} H^{\frac{3}{n}}\,.
\ee

It is interesting to notice that, for $n=3$, the explicit dependence from the reduced Planck mass $M_P$ drops and we recover the relation $m_{3/2}> H$ found in \cite{Kallosh:2004yh} and \cite{Kolb:2021xfn,Kolb:2021nob}. However, the origin of the bound in this context is very different. It is related to a number of light species of the UV theory  which enter the effective theory and then decrease the quantum gravity cut-off. When $n=3$, as we have shown in the previous section, we have no EFT of supersymmetry breaking as $m_{3/2}\simeq\Lambda_{QG}$.

In the following sub-sections, we will provide a number of concrete phenomenological applications in quasi-de Sitter background.

\subsection{A lower bound on $m_{3/2}$ from CMB}

The inflationary paradigm provides a concrete setting to consider the implications of the bounds presented above. According to this scenario, today Cosmic Microwave Background (CMB) experiments are probing scales, which crossed the Hubble horizon around 60 e-foldings before the end of inflation. The latest CMB measurements \cite{Akrami:2018odb} have set a bound on the separation between the Hubble inflationary energy and the Planck scale, namely

\be\label{Planck}
H<2.5 \cdot 10^{-5}M_P\,.
\ee
This is in fact directly related to the observational upper bound on tensor-to-scalar ratio $r\lesssim 0.06$. In the slow-roll approximation, the relation between $H$ and $r$ is given by
\be\label{Hr}
H=\sqrt{\frac{\pi^2 A_s\ r}{2}} M_P\simeq 10^{-4} \sqrt{r}\  M_P\,,
\ee
where $A_s$ is the amplitude of scalar perturbations and all the quantities are meant as calculated at the scales which exited the Hubble horizon around 60 e-foldings before the end of inflation.

The bound eq.~\eqref{boundHm32}, provided by the GMC on the gravitino mass in terms of the Hubble parameter, can be therefore recast in terms of the tensor-to-scalar ratio (calculated at horizon exit) as
\be\label{m32r}
m_{3/2}> \left(10^{-12}\ r^{\frac{3}{2}}\right)^{\frac{1}{n}}\ M_P\,,
\ee
where we remind that $n$ gives the scaling of the mass $m$ of the infinite tower in terms of $m_{3/2}$ as in eq.~\eqref{massm32Mp}.

We can use this bound \eqref{m32r} in order to get information on the gravitino mass range allowed by the GMC, if we will detect primordial gravitational waves. In this case, our assumption is that the gravitino mass does not change much from the time of inflation up to now (this typically happens, for example, if the uplifting mechanism to realize inflation in string theory is of the form of what discussed in Appendix \ref{AppendixUPdS}).

Next generation CMB experiments \cite{Abazajian:2016yjj,Ade:2018sbj} (such as CMB-S4 and the Simons Observatory) have a target of $r=\mathcal{O}(10^{-3})$.
This means that, in the case of the strong GMC ($n=1$), any detection of B-modes in the next future would point at a lower bound for the gravitino mass of the order
\be\label{m32GeV}
m_{3/2}\gtrsim  10^2\ \text{GeV}\,.
\ee

One could also consider the highest inflationary energy today compatible with observations (i.e. the value that saturates the bound eq.~\eqref{Planck}). The eventuality of such a detection would of course correspond to a much stronger bound of the order

\be\label{m32TeV}
m_{3/2}\gtrsim 10^2\ \text{TeV}\,.
\ee

\begin{figure}[t!]
	\begin{center}
		\includegraphics[scale=1]{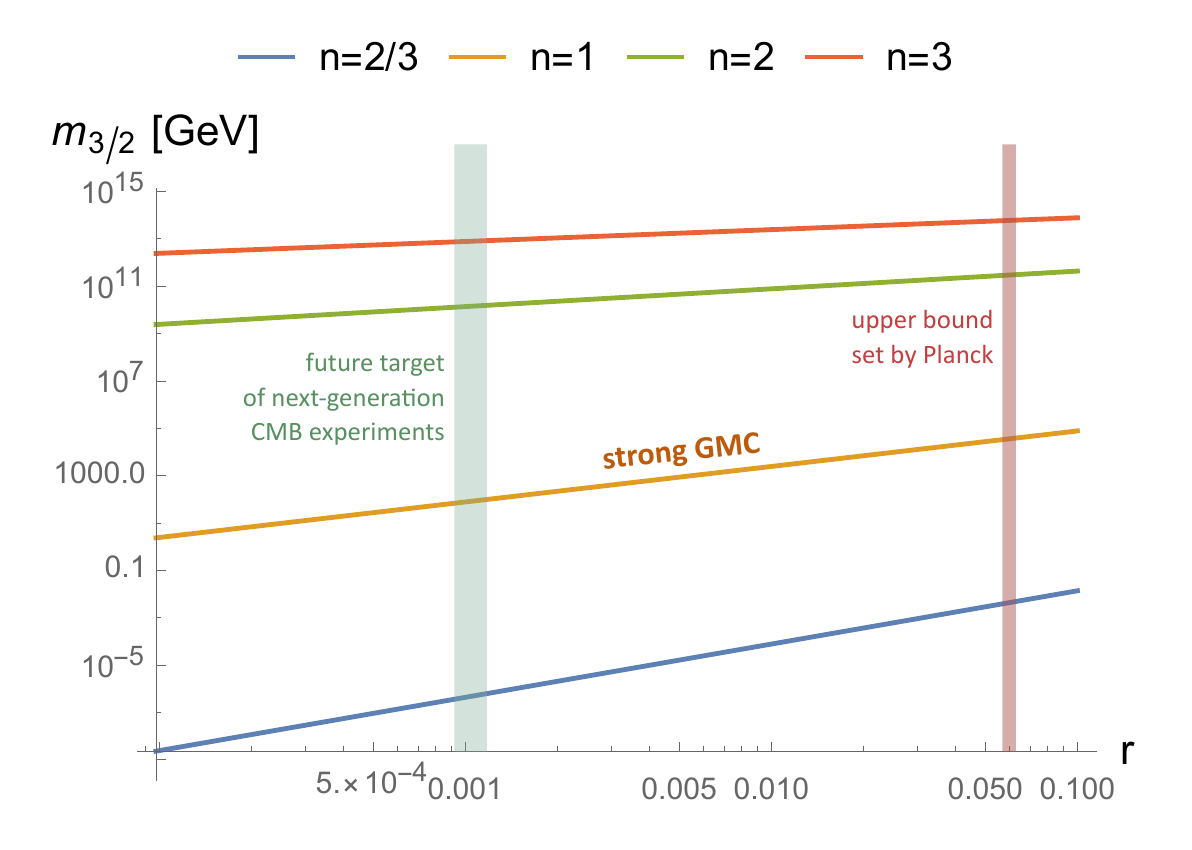}
		\vspace*{-0.2cm}\caption{Log-log plot of the lower bounds on the gravitino mass (in GeV) in terms of the tensor-to-scalar ratio measured at CMB scales, as given by eq.~\eqref{m32r}. The plot is given for $n=\{2/3, 1,2,3\}$ and the orange line corresponds to the predictions of the strong GMC. The red vertical line shows the upper bound $r\lesssim 0.06$ set by Planck \cite{Akrami:2018odb} while the green vertical area corresponds to the future target $r=\mathcal{O}(10^{-3})$ of next-generation experiments \cite{Abazajian:2016yjj,Ade:2018sbj}.}
		\label{m32rplot}
	\end{center}
\end{figure}

In fig.~\ref{m32rplot}, we show the lower bound \eqref{m32r} for different values of $n$ in logarithmic scale. One can also appreciate the estimates extracted above, eq.~\eqref{m32GeV} and eq.~\eqref{m32TeV}, being at the intersection between the orange line ($n=1$) and the vertical green and red lines.

\subsection{An upper bound on the scalar field range in terms of $m_{3/2}$}

In a quasi-de Sitter phase sustained by the energy of a scalar potential one has typically a field displacement $\Delta\phi$. Inflation provides a primary example of this situation.  The SDC \cite{Ooguri:2006in} implies that while traversing such a distance, an infinite tower of states will have decreasing masses and the quantum gravity cut-off will decrease as \cite{Scalisi:2018eaz}
\be
\Lambda_{QG}= \Lambda_0\ e^{-\lambda \Delta\phi}\,,
\ee
where $\Lambda_0\leq M_P$ is the original naive cut-off of the EFT. This implies a universal upper bound on the scalar field variation such as
\be
\Delta\phi<\frac{1}{\lambda}\log\frac{M_P}{\Lambda_{QG}}\,.
\ee
If we now use the expression for the cut-off eq.~\eqref{QGcutoff}, we obtain
\be\label{Deltaphim32}
\Delta\phi<\frac{n}{3\lambda}\log\frac{M_P}{m_{3/2}}\,,
\ee
 which provides a model independent upper bound of the scalar field variation in terms of the gravitino mass. We remind that $n$ defines the power-dependence of the mass tower in terms of $m_{3/2}$ as given by eq.~\eqref{massm32Mp}. Note that, to use the formula of the cut-off eq.~\eqref{QGcutoff} in terms of $m_{3/2}$, we are assuming that $\Delta\phi$ is the distance along which the gravitino mass decreases. We therefore identify implicitly the tower predicted by the GMC with the tower of the SDC. If we set $n\simeq\lambda\simeq1$,\footnote{Lower bounds on the exponential rate $\lambda$ have been proposed for example in \cite{Andriot:2020lea,Gendler:2020dfp}.} we notice that the bound eq.~\eqref{Deltaphim32} would constrain large scalar field variation ({\it i.e.} $\Delta\phi>1$) just for very high values of the gravitino mass quite close to the Planck scale, namely for $m_{3/2}>10^{-2} M_P$.

\subsection{Gravitino coupling constant and Hubble parameter}

In a background with a positive cosmological constant, the work \cite{Montero:2019ekk} has shown that there exists a general lower bound on the mass of a charged particle. In the case of the gravitino, this reads
\be
m_{3/2}>\sqrt{q_{3/2}\ g_{3/2}\ M_P H}\,.
\ee
If we combine this with the first inequality of eq.~\eqref{WGCbound} (which holds independently from the specific background), we obtain a bound between the Hubble parameter and the gravitino gauge coupling constant of the form
\be
\frac{g_{3/2}^5}{q_{3/2}}>\frac{H}{M_P}\,.
\ee
If we assume that $g_{3/2}<1$, the latter represents a stronger bound than the one obtained from the magnetic WGC and requiring perturbative control of our EFT, that is \mbox{$H<\Lambda_{QG}<g_{3/2}\ M_P$}. Detection of B-modes in next-generation CMB experiments would point at a bound $g_{3/2}\gtrsim 0.08$.

\section{Conclusions}\label{sum}

The focus of this work has been the role of the gravitino, in the context of the swampland program. We proposed a new swampland criterion, the Gravitino Mass Conjecture (GMC), stating that the limit of vanishing gravitino mass leads to the breakdown of the effective description. Our claim is supported by large classes of models arising from string compactifications to four dimensions, with $\mathcal{N}=1,2$ supersymmetry on a generic background. 
Note however that the GMC however does not forbid string vacua, which are strictly supersymmetric and have a vanishing gravitino mass $m_{3/2}=0$. The analogy with the Higuchi bound \cite{Higuchi:1986py} for the graviton (although this is valid just in de Sitter backgrounds) seems natural: a massless graviton is certainly allowed but arbitrary small values of the mass are not.

We discussed extensively the connection of the GMC to other existing swampland conjectures. In particular, we pointed out similarities and differences of the GMC with respect to the Anti-de Sitter Distance Conjecture (ADC) \cite{Lust:2019zwm}. In the case of supersymmetric AdS vacua, where $m_{3/2}^2$ and $\Lambda$ can be simply identified as given by eq.~\eqref{m32SUSYAdS}, the two conjectures are equivalent to each other. Furthermore, given the general supergravity lower bound eq.~\eqref{vbound} on the gravitino mass in terms of the value of the scalar potential, one has also that, for AdS spaces, the GMC implies always the ADC: the limit of small gravitino mass necessarily implies a small value of the potential. More generally, this implication (\textit{i.e. }$m_{3/2}\rightarrow 0$ implies $|V|\rightarrow 0$) can be extended also to positive potentials when the gravitino mass goes to zero in the limit of large internal volume, that is when $e^K$ goes to zero. In fact, the factor $e^K$ appears overall both in the definition of the gravitino mass eq.~\eqref{m32def} and in the expression of the scalar potential. However, the opposite  (\textit{i.e. }$|V|\rightarrow 0$ implies $m_{3/2}\rightarrow 0$) is in general not true. One may keep a finite value of the gravitino mass while going to the limit of vanishing potential via compensation of supersymmetry breaking F- and/or D-terms. In this case, the ADC and the GMC would lead to different conclusions concerning the tower behaviour and the associated phenomenology . This applies specifically to non-supersymmetric vacua, especially in cases for which $m_{3/2}$ and the cosmological constant $\Lambda$ are  separated from each other. 
It is therefore an interesting question, if such a {\it decoupling} between $m_{3/2}$ and $\Lambda$ is possible in microscopic string constructions, with a single tower associated to the gravitino mass or possibly to two distinct towers. Instead, if such a decoupling belongs to the swampland, this would mean that $m_{3/2}$ and $\Lambda$ are always parametrically linked together. They may, for example, satisfy a power relation such as
\be\label{m32Lambdapower}
m_{3/2}^2\sim |\Lambda|^p
\ee
in Planck units, which for $p=1$ corresponds to saturating the lower bound eq.~\eqref{vbound}, at least for AdS vacua. Therefore, in the case of a parametric link between $m_{3/2}$ and $\Lambda$, the ADC and GMC would be even more closely related (equivalent just in the case of $p=1$). We would like however to notice that, assuming for example $m_{3/2}^2\sim |\Lambda|$, this would enforce also a further fine-tuning of the supersymmetry breaking terms, which should follow a relation such as $V_F+V_D\sim m_{3/2}^2\sim |\Lambda|$. Moreover, applying this relation to the case of the current cosmic acceleration with $\Lambda\sim 10^{-120}\ M_P$, we obtain that the mass of the gravitino should be $m_{3/2} \sim 10^{-33}\ \text{eV}$ and a consequent supersymmetry breaking scale $M_{SUSY}\sim 10^{-15}\ \text{TeV}$, which seems in contradiction with current observations. Perhaps, if one would like to stick to this link as given by eq.~\eqref{m32Lambdapower}, one may use current observational bounds on the scale up to which we have not observed SUSY and get a constraint on the power $p$.

In the framework of $\mathcal{N}=2$ supergravity, we identified the existence of a relation between the gravitino mass and the abelian gauge coupling, allowing us to interpret the GMC in view of the absence of global symmetries in quantum gravity. 

After inspecting the quantum gravity cut-off predicted by the GMC, we discussed also phenomenological implications of our conjecture on backgrounds with positive energy. Specifically, we showed that the GMC predicts a universal lower bound eq.~\eqref{boundHm32} for the mass of the gravitino in terms of the Hubble parameter, under which the EFT breaks down. By employing this, we can show that any detection of primordial gravitational by next generation CMB experiments \cite{Abazajian:2016yjj,Ade:2018sbj} would suggest a lowest possible value of around 10 GeV for the mass of the gravitino. 

Our work can be extended in various directions and we would like to sketch some of them below. The proposed GMC is a statement on the obstruction in taking the limit of a vanishing mass. In this respect, as we already highlighted,  it has some similarities with the Higuchi bound on the mass of particles of integer higher spin \cite{Higuchi:1986py}. It would be interesting to understand if it is possible to re-derive such a bound exclusively from swampland considerations. 

Another possible development would be to investigate further the relation between the gravitino mass and gauge coupling derived in $\mathcal{N}=2$ supergravity and understand under which conditions such a relation survives in models with minimal supersymmetry. This could shed light on the origin of the $\mathcal{N}=1$ scalar potential from higher supersymmetric models. At the same time, one could also try to understand if swampland ideas are strong enough to fix its functional form.
\vspace{0.5cm}

{\bf Note added in v2}: two days after the present work appeared on arXiv, the paper \cite{Castellano:2021yye} was posted too, proposing the same conjecture on the gravitino mass. The two works have therefore some overlap but also complementary discussions.

\vskip0.5cm
\vspace{10px}
{\bf Acknowledgements}
\vskip0.1cm
\noindent

We thank M.~Cicoli, G.~Dall'Agata, G.~Dvali, F.~Farakos, A.~Hebecker, E.~McDonough, E.~Palti, C.~Vafa, T.~Weigand and Y.~M.~Zinoviev for very useful discussions. The work of N.C.~is supported by an FWF grant with the number P 30265. The work of D.L.~is supported  by the Origins Excellence Cluster.

\appendix

\section{Position of the de Sitter vacuum after uplifting}\label{AppendixUPdS}

In this appendix, we provide a simple derivation of the fact that the position $\mathcal{V}_1$  of the de Sitter minimum will be almost unchanged with respect to the position $\mathcal{V}_0$ of the anti-de Sitter vacuum, if the uplifting contribution scales as a negative power of the internal volume, that is $V_{up}=c/\mathcal{V}^p$ (with $c$ and $p$ positive). We want to show that $\delta\mathcal{V}=\mathcal{V}_1-\mathcal{V}_0\ll1$.

We consider an expansion of the potential $V_{AdS}$ around its minimum, prior to the uplift:
\be\label{VAdSexpansion}
V_{AdS}(\mathcal{V})= V_{AdS}(\mathcal{V}_0) + b (\mathcal{V}-\mathcal{V}_0)^2\,,
\ee
where the first term gives the depth of anti-de Sitter vacuum and $b=1/2V_{AdS}''(\mathcal{V}_0)>0$, with the prime indicating derivative with respect to the volume. The minimum of the full potential $V_{dS}=V_{AdS}+V_{up}$ is such that
\be
V_{dS}'|_{\mathcal{V}=\mathcal{V}_1}= (V_{AdS}'+V_{up}')|_{\mathcal{V}=\mathcal{V}_1}=0\,.
\ee
After including the expansion \eqref{VAdSexpansion} and the explicit form of $V_{up}$, we have
\be
\left[ b (\mathcal{V}-\mathcal{V}_0)^2 + \frac{c}{\mathcal{V}^p}\right]'\big|_{\mathcal{V}=\mathcal{V}_1}=0\,,
\ee
which gives
\be
\delta\mathcal{V}=\mathcal{V}_1-\mathcal{V}_0= \frac{cp}{2b}\frac{1}{\mathcal{V}_1^{p+1}}\,.
\ee
At large volume we have $\delta\mathcal{V}\ll1$ and the shift of the minimum after uplifting is thus negligible.

\section{Rewriting the $\mathcal{N}=2$ scalar potential}\label{appB}

In this appendix, we give some details on how to rewrite the scalar potential of $\mathcal{N}=2$ supergravity in $\mathcal{N}=1$ languange. We concentrate on each term in \eqref{VN=2} separately.

Let us start from $V_1$. This term depends entirely on the vector multiplets. Using the special geometry relation
 \begin{equation}
 (\partial_i \mathcal{P}^0_\Lambda) L^\Lambda = - \mathcal{P}^0_\Lambda f_i^\Lambda, 
 \end{equation}
which can be derived from $\mathcal{P}^0_\Lambda L^\Lambda=0$, we have
 \begin{equation}
 \begin{aligned}
 \label{V1app}
 {V}_1 &= g_{i \bar \jmath} k^i_\Lambda k^{\bar \jmath}_\Sigma \bar L^\Lambda L^\Sigma =g^{i \bar \jmath}\partial_i \mathcal{P}^0_\Lambda \partial_{\bar \jmath}\mathcal{P}^0_\Sigma \bar L^\Lambda L^\Sigma= U^{ \Sigma \Lambda} \mathcal{P}^0_\Lambda \mathcal{P}^0_\Sigma=-\frac12 ({\rm Im}\, \mathcal{N}^{-1})^{\Lambda \Sigma} \mathcal{P}^0_\Lambda \mathcal{P}^0_\Sigma.
 \end{aligned}
 \end{equation}
Notice that the minus sign is consistent with the fact that the matrix ${\rm Im } \mathcal{N}_{\Lambda \Sigma}$ is negative definite.
 
 Next we consider $V_2$. This term is more involved since it contains the metric $h_{uv}$ and thus information about the quaternionic nature of the manifold. By using that 
 \begin{equation}
\mathcal{P}^x_\Lambda \mathcal{P}^x_\Sigma = \frac12 \left( \mathcal{P}_\Lambda \bar {\mathcal{P}}_\Sigma + \mathcal{P}_\Sigma \bar {\mathcal{P}}_\Lambda \right) + \mathcal{P}_\Lambda^3 \mathcal{P}_\Sigma^3,
\end{equation}
where $\mathcal{P}_\Lambda = \mathcal{P}_\Lambda^1 - i  \mathcal{P}_\Lambda^2$, we have
  \begin{equation}
 \begin{aligned}
  h_{uv} k^u_\Lambda k^v_\Sigma \bar L^\Lambda L^\Sigma&= \frac{1}{12} h^{uv}\nabla_u \mathcal{P}^x_\Lambda \nabla_v \mathcal{P}^x_\Sigma \bar L^\Lambda L^\Sigma\\
 &= \frac{1}{12} h^{uv} \bar L^\Lambda L^\Sigma \left[\nabla_u\bar{ \mathcal{P}}_{(\Lambda}\nabla_{|v|}{ \mathcal{P}}_{\Sigma)} + \nabla_u \mathcal{P}_\Lambda^3 \nabla_v \mathcal{P}^3_\Sigma\right]\\
 &= \frac{1}{12} h^{uv} \bar L^\Lambda L^\Sigma \left[\nabla_u\bar{ \mathcal{P}}_{\Lambda}\nabla_v{ \mathcal{P}}_{\Sigma}- \nabla_u\bar{ \mathcal{P}}_{[\Lambda}\nabla_{|v|}{ \mathcal{P}}_{\Sigma]}+ \nabla_u \mathcal{P}_\Lambda^3 \nabla_v \mathcal{P}^3_\Sigma\right].
\end{aligned}
 \end{equation}
 Then, one uses
 \begin{equation}
 h^{uv} \nabla_u \mathcal{P}_{[\Lambda}\nabla_{|v|} \bar{\mathcal{P}}_{\Sigma]} = 4 \mathcal{P}_{[\Lambda} \bar{\mathcal{P}}_{\Sigma]} + 4 i f^\Gamma_{\Lambda \Sigma} \mathcal{P}^3_\Gamma,
 \end{equation}
 which can be derived from the quaternionic equivariance condition, together with the special geometry relation 
  \begin{equation}
 \bar L^\Lambda L^\Sigma f_{\Lambda \Sigma}^\Gamma = \bar L^\Lambda k^i_\Lambda f_i^\Gamma = \frac i2 ({\rm Im} \mathcal{N}^{-1})^{\Gamma \Lambda} \mathcal{P}_\Lambda^0\, ,
 \end{equation}
 to get
 \begin{equation}
  \begin{aligned}
  \label{hkk}
    h_{uv} k^u_\Lambda k^v_\Sigma \bar L^\Lambda L^\Sigma&= \frac{1}{12} h^{uv} \bar L^\Lambda L^\Sigma (\nabla_v \bar{\mathcal{P}}_\Lambda \nabla_b \mathcal{P}_\Sigma + \nabla_u \mathcal{P}^3_\Lambda \nabla_v \mathcal{P}^3_\Sigma)\\
    &-\frac13 \bar L^\Lambda L^\Sigma \mathcal{P}_{[\Lambda}\bar{\mathcal{P}}_{\Sigma]} - \frac16 ({\rm Im}\mathcal{N}^{-1})^{\Lambda \Sigma}\mathcal{P}_\Lambda^3\mathcal{P}_\Sigma^0.
 \end{aligned}
 \end{equation}
 We notice that the second line contains mixed terms $\mathcal{P}_{[\Lambda} \bar{\mathcal{P}}_{\Sigma]}$ and $\mathcal{P}_\Lambda^3 \mathcal{P}_\Sigma^0$.  The latter are needed to reconstruct the mixed product in $V_D$, equation \eqref{VD2}, but the coefficient is not the correct one. However, if we write
 \begin{equation}
 V_2 = 4 h_{uv} k^u_\Lambda k^v_\Sigma \bar L^\Lambda L^\Sigma = (6-2)h_{uv} k^u_\Lambda k^v_\Sigma \bar L^\Lambda L^\Sigma
 \end{equation}
 and we substitute \eqref{hkk} only in the first factor, namely the one with coefficient 6, the term $\mathcal{P}_\Lambda^3 \mathcal{P}_\Sigma^0$ thus generated will have the correct coefficient to provide the mixed product  in $V_D$, while the term $\mathcal{P}_{[\Lambda} \bar{\mathcal{P}}_{\Sigma]}$ will cancel precisely against an analogous one coming from $V_3$. Therefore, we recast $V_2$ into the form
 \begin{equation}
 \label{V2app}
\begin{aligned}
{V}_2 &= -2h_{uv} k^u_\Lambda k^v_\Sigma\\
&+\left[ \frac12 h^{uv} \bar L^\Lambda L^\Sigma (\nabla_v \bar{\mathcal{P}}_\Lambda \nabla_b \mathcal{P}_\Sigma + \nabla_u \mathcal{P}^3_\Lambda \nabla_v \mathcal{P}^3_\Sigma)-2 \bar L^\Lambda L^\Sigma \mathcal{P}_{[\Lambda}\bar{\mathcal{P}}_{\Sigma]} -  ({\rm Im}\mathcal{N}^{-1})^{\Lambda \Sigma}\mathcal{P}_\Lambda^3\mathcal{P}_\Sigma^0\right].
\end{aligned} 
\end{equation}

Finally we look at $V_3$, for which we have
 \begin{equation}
 \label{V3app}
\begin{aligned}
{V}_3 &= (U^{\Lambda \Sigma} - 3 \bar L^\Lambda L^\Sigma) \mathcal{P}^x_{\Lambda} \mathcal{P}^x_\Sigma=(U^{\Lambda \Sigma} - 3 \bar L^\Lambda L^\Sigma)  \left( \mathcal{P}_{(\Lambda} \bar {\mathcal{P}}_{\Sigma)}  + \mathcal{P}_\Lambda^3 \mathcal{P}_\Sigma^3 \right)\\
&= U^{\Lambda \Sigma} \mathcal{P}_\Lambda \bar{\mathcal{P}}_\Sigma-3 \bar L^\Lambda L^\Sigma \bar{\mathcal{P}}_\Lambda \mathcal{P}_\Sigma +(U^{\Lambda \Sigma} - 3 \bar L^\Lambda L^\Sigma) \mathcal{P}^3_\Lambda \mathcal{P}^3_\Sigma +2 \bar L^\Lambda L^\Sigma \bar{\mathcal{P}}_{[\Lambda}\mathcal{P}_{\Sigma]}\\
&= g^{i  \bar \jmath} D_i \mathcal{W} D_{\bar \jmath} \bar{\mathcal{W}} - 3 \mathcal{W} \bar {\mathcal{W}} + 2 \bar L^\Lambda L^\Sigma  \bar{\mathcal{P}}_{[\Lambda}\mathcal{P}_{\Sigma]} - \frac12 ({\rm Im}\mathcal{N}^{-1})^{\Lambda \Sigma} \mathcal{P}_\Lambda^3 \mathcal{P}_\Sigma^3 - 4 \bar L^\Lambda L^\Sigma  {\mathcal{P}}_{\Lambda}^3\mathcal{P}_{\Sigma}^3.
\end{aligned}
\end{equation}
 
 By adding up \eqref{V1app}, \eqref{V2app} and \eqref{V3app}, one obtains \eqref{V2final}.


\end{document}